\documentclass{article}
\usepackage{graphicx}
\usepackage{amsmath}
\usepackage{harvard}

\input{tcilatex}

\begin{document}

\title{Bell's theorem without inequalities and only two distant observers}
\author{P.K.Aravind \\
Physics Department, Worcester Polytechnic Institute,\\
Worcester, MA 01609.}
\date{\today}
\maketitle

\begin{abstract}
A proof of Bell's theorem without inequalities and involving only two
observers is given by suitably extending a proof of the Bell-Kochen-Specker
theorem due to Mermin. This proof is generalized to obtain an
inequality-free proof of Bell's theorem for a set of $n$ Bell states (with $%
n $ odd) shared between two distant observers. A generalized CHSH inequality
is formulated for $n$ Bell states shared symmetrically between two observers
and it is shown that quantum mechanics violates this inequality by an amount
that grows exponentially with increasing $n$.
\end{abstract}

In two recent papers[1,2], Cabello gave a proof of Bell's theorem without
inequalities by using a special state of four qubits shared between two
distant observers. This improved upon the classic proof of Greenberger,
Horne and Zeilinger[3] and Mermin[4] by reducing the number of distant
observers from three to two. The purpose of this paper is to describe a
variant of Cabello's proof that avoids one of its shortcomings and can also
be generalized to apply to a suitable entangled state of $2n$ qubits
(namely, $n$ identical Bell states) shared symmetrically between two
observers. The present proof, like Cabello's[2] and several others before
it[5], uses a common framework to prove both the Bell-Kochen-Specker
(BKS)[6] and Bell[7] theorems. However, while Cabello proceeds backwards
from the stronger (Bell) to the weaker (BKS) theorem, we proceed in the
opposite direction. Our approach has the advantage over Cabello's that it
makes no use of either entanglement or communication between observers in
proving the BKS theorem, and invokes these additional elements only in
passing from the BKS to the Bell theorem.

The present proof is similar in overall structure to an earlier proof by
Heywood and Redhead[8], although it differs in several specific respects.
Both proofs exploit EPR type correlations to derive non-contextuality from
locality, and then use a Kochen-Specker argument to establish the
inevitability of non-locality. However while Heywood and Redhead use a
singlet state of two spin-1 particles and the original Kochen-Specker
argument in carrying out their proof, we use $n$ Bell states shared
symmetrically between two observers and a more transparent variant of the
Kochen-Specker argument due to Mermin[9] in carrying out our proof. We
believe the latter route offers both a gain in simplicity relative to the
Heywood-Redhead argument and also has a more direct contact with experiment,
since only qubits are involved in the present scheme and the question of
imperfect correlations and/or detector efficiencies is also addressed.

Figure 1 shows a 3 x 3 array of observables pertaining to a pair of qubits
used by Mermin[9] to prove the BKS theorem. Mermin's proof is based on the
elementary observations that: (a) each observable has only the eigenvalues $%
\pm 1$, (b) the observables in any row or column of the array form a
mutually commuting set, and (c) the product of the observables (and hence
their eigenvalues) in any row or column is +1, with the exception of the
last column for which this product is --1. Armed with these facts, Mermin's
argument proceeds as follows:

Suppose an experimenter, Alice, who has two qubits in her possession carries
out the measurements corresponding to the commuting observables in one of
the rows or columns of Mermin's square. The result will be a set of\ \ +1s
and --1s for the measured eigenvalues satisfying the product constraint
mentioned earlier. The product constraint can be restated as the ''sum''
constraint that the total number of -1s for any triad of commuting
observables is always even, except for the last column for which it is odd.
Now if Alice is a ``realist'' and believes that the eigenvalues she measures
merely reflect preexisting properties of the qubits, she would be tempted to
assign the value +1 or --1 to each of the nine observables in Mermin's
square in such a way that all the sum constraints on their values are met.
However this is easily seen to be impossible by counting the total number of
-1s in the square in two different ways: firstly, by summing over the rows
(which leads to an even number) and, secondly, by summing over the columns
(which leads to an odd number). This contradiction shows the impossibility
of assigning preexisting values (or ''elements of reality''[10]) to the
concerned observables and constitutes Mermin's proof of the BKS theorem.

However the above BKS proof has the objectionable feature that an observable
is assigned the same value whether it is measured as part of a row or column
of observables. This assumption of ``noncontextuality'' has no empirical
basis and, in the opinion of many physicists (including Bell himself[6]),
considerably diminishes the force of the BKS theorem[11]. We now show how to
rectify this defect and thereby promote Mermin's BKS proof into a proof of
Bell's theorem. To do this we enlist the help of a second experimenter, Bob,
give him two qubits of his own, and allow him to do everything Alice can at
a location far removed from hers. The trick to ensuring that Alice and Bob
can jointly prove Bell's theorem is that the four qubits given to them are
in the entangled state

\begin{equation}
\left| \Psi \right\rangle =\frac{1}{\sqrt{2}}(\left| 00\right\rangle +\left|
11\right\rangle )_{13}\otimes \frac{1}{\sqrt{2}}(\left| 00\right\rangle
+\left| 11\right\rangle )_{24}=\frac{1}{2}\left( \left| 0000\right\rangle
+\left| 0101\right\rangle +\left| 1010\right\rangle +\left|
1111\right\rangle \right) ,
\end{equation}
where $0$ and $1$ denote basis states of a qubit and the subscripts $1,..,4$
in the middle expression indicate the relative positions of these qubits in
the last, expanded form of the state $\left| \Psi \right\rangle $. In other
words, $\left| \Psi \right\rangle $ consists of a pair of identical Bell
states, with one member of each pair (qubits $1$ and $2$) going to Alice and
the others (qubits $3$ and $4$) going to Bob. It is also assumed that the
first and second members of each two particle observable in Fig.1 refer to
qubits $1$ and $2$ for Alice and qubits $3$ and $4$ for Bob.

The state (1) possesses the interesting property that $O_{i}^{A}O_{i}^{B}%
\left| \Psi \right\rangle =\left| \Psi \right\rangle $ for $i=1,...,9,$
where $O_{i}^{A}$ is any one of Alice's nine observables and $O_{i}^{B}$ is
the same observable for Bob. This property implies that if Alice and Bob
measure identical observables on their qubits, they always obtain the same
eigenvalues, even if their measurements are carried out at spacelike
separations. These correlated outcomes suffice to establish that all of
Alice's observables are ''elements of reality'', and that the same is true
of Bob's. For Bob (or Alice) can use his (or her) measurement of a
particular observable to instantly predict the outcome of the other person's
measurement of the same observable at a distant location without disturbing
that person's qubits in any way. Note further that either person's ability
to predict the value of the other's observable is independent of whether the
latter is measured by itself or as part of any commuting triad it happens to
be a member of[12]. But this last statement is just the assumption of
noncontextuality, now justified on the basis of the correlations in state
(1) and the principle of locality, and serves to promote Mermin's earlier
BKS proof \ into a full fledged proof of Bell's theorem.

Cabello's proof[2] differs from ours in that state (1) is replaced by a
direct product of singlets and the nine observables measured by Alice and
Bob are not identical. However a more significant difference is that, in
Cabello's scheme, Alice and Bob are required to collaborate in measuring
five non-local observables each made up of their separate observables. While
the measurement of these non-local observables poses no problems for a Bell
test, it imposes the unnecessary burden on a BKS test of requiring
communication between the observers to achieve its goals.

The present BKS-Bell proof suggests a joint laboratory experiment for
verifying the BKS and Bell theorems. However its practical realization is
complicated by the fact that each observer needs to be able to measure a
sequence of three commuting two-particle observables on his/her qubits. Such
''non-demolition'' measurements are possible to envision in principle[13],
but they are rather challenging to carry out in practice.

The above Bell proof can be generalized to a set of $n$ Bell states (with $n$
odd) shared symmetrically between two observers. Consider the following $%
(n+2)$ sets of mutually commuting $n$-qubit observables, where each
commuting set is shown on a separate line and the superscripts on the Pauli
operators refer to the different qubits:

\begin{eqnarray}
&&\sigma _{x}^{1}\sigma _{z}^{2}\sigma _{x}^{3},\sigma _{x}^{2}\sigma
_{z}^{3}\sigma _{x}^{4},...,\sigma _{x}^{i}\sigma _{z}^{i+1}\sigma
_{x}^{i+2},...,\sigma _{x}^{n-1}\sigma _{z}^{n}\sigma _{x}^{1},\sigma
_{x}^{n}\sigma _{z}^{1}\sigma _{x}^{2},\sigma _{z}^{1}\sigma _{z}^{2}\cdot
\cdot \cdot \sigma _{z}^{n}, \\
&&\sigma _{x}^{1},\sigma _{z}^{2},\sigma _{x}^{3},\sigma _{x}^{1}\sigma
_{z}^{2}\sigma _{x}^{3} \\
&&\sigma _{x}^{2},\sigma _{z}^{3},\sigma _{x}^{4},\sigma _{x}^{2}\sigma
_{z}^{3}\sigma _{x}^{4} \\
&&...\text{ \ \ }...\text{ \ \ }...\text{ \ \ }...\text{ \ \ }...\text{ \ \ }%
... \\
&&\sigma _{x}^{n},\sigma _{z}^{1},\sigma _{x}^{2},\sigma _{x}^{n}\sigma
_{z}^{1}\sigma _{x}^{2} \\
&&\sigma _{z}^{1},\sigma _{z}^{2},\cdot ,\cdot ,\cdot ,\sigma
_{z}^{n},\sigma _{z}^{1}\sigma _{z}^{2}\cdot \cdot \cdot \sigma _{z}^{n}.
\end{eqnarray}
Each line after the first consists of one of the observables in the first
line together with all the single particle observables of which it is made
up. There are $(n+1)+2n=3n+1$ distinct observables in all, each of which
occurs in exactly two commuting sets. The observables have the further
properties that: (a) each has only the eigenvalues $\pm 1$, and (b) the
product of the observables (and hence their eigenvalues) in any commuting
set is $+1,$ with the exception of the first set for which this product is $%
-1$.

A BKS proof can now be constructed as follows. Suppose Alice is given $n$
qubits and allowed to measure the above observables on them. If Alice is a
''realist'' and believes that the eigenvalues she measures already preexist
in the qubits, she would be tempted to assign the value $+1$ or $-1$ to each
of the observables in such a way that the products of the values
corresponding to each of the rows in (2)-(7) is $+1$, with the exception of
the first row for which it is $-1$. However this assignment is easily seen
to be impossible by taking the product of all the value equations, for one
then finds that the product of all the left sides is $+1$ (because each
observable value occurs exactly twice) whereas the product of all the right
sides is $-1$ (because of the first equation in this chain). This
contradiction shows the impossibility of\ assigning preexisting values to
the observables and proves the BKS theorem for a system of $n$ qubits. We
now show how to justify the assumption of noncontextuality made in this
argument and thus promote it into a proof of Bell's theorem.

Consider the tensor product of $n$ Bell states $\frac{1}{\sqrt{2}}(\left|
00\right\rangle +\left| 11\right\rangle )_{11^{^{\prime }}}\otimes \frac{1}{%
\sqrt{2}}(\left| 00\right\rangle +\left| 11\right\rangle )_{22^{^{\prime
}}}\otimes ...\otimes \frac{1}{\sqrt{2}}(00+11)_{nn^{^{\prime }}}$, and
suppose that the unprimed member of each Bell state is given to Alice and
the primed member to Bob. Suppose that Alice and Bob are each allowed to
measure any of the $3n+1$ observables in (2)-(7) on their respective qubits.
Then it is not difficult to verify that if Alice and Bob measure identical
observables on their qubits, they always obtain the same eigenvalues. From
this perfect correlation one can argue, as before, that either observer's
observables are elements of reality and hence that the BKS proof based on
(2)-(7) can be promoted into a Bell proof.

The $n=3$ case of our BKS proof was given by Mermin[9], who arranged the ten
relevant observables[14] at the vertices of a pentagram in such a way that
each set of commuting observables lay along one of its edges. Mermin then
converted this BKS proof into a Bell proof by assuming that the three qubits
were in a GHZ state. The $n=5$ case of our BKS proof was given by DiVincenzo
and Peres[15], who pointed out that it could be promoted into a Bell proof
if the five qubits were assumed to be in a state corresponding to one of the
logical codewords of the five-qubit single error-correcting code discussed
in [16]. The foregoing Bell proofs of Mermin and of DiVincenzo and Peres
involve three and five separated observers, respectively, whereas our proofs
involve only two observers who however share three or five Bell states
symmetrically between themselves.

In an interesting paper dealing with the transition from quantum nonlocality
to classical behavior in the limit of an infinite number of particles,
Pagonis et al[17] introduce several alternative sets of commuting
observables for $N$ qubits that allow closely related proofs of the BKS and
Bell theorems to be given. Their Bell proofs require distributing a
simultaneous eigenstate of the commuting observables to $N$ observers who
then carry out measurements on their individual qubits in a generalization
of the GHZ protocol. It is worth noting that the observables proposed by
Pagonis et al can also be used by just two observers to carry out a joint
BKS-Bell proof provided that they share $N$ Bell states symmetrically among
themselves and exploit the identity of the eigenvalues of similar
observables measured at their two ends. An interesting difference between
the observables in (2)-(7) and those proposed by Pagonis et al is that our
observables generally involve no more than three qubits whereas those of
Pagonis et al always involve all $N$ qubits simultaneously. Another, more
significant, difference is that in the scheme of Pagonis et al (and also
that of Mermin-GHZ), the observables that are used to prove the BKS theorem
are closely related to the entangled state used in the\ later Bell proof
(the latter being a simultaneous eigenstate of the nontrivial commuting
observables in the BKS proof). In contrast to this, the observables used in
our BKS proofs bear no relation to the entangled (Bell) states used in the
later Bell proofs.

Our Bell proofs assume that Alice and Bob share perfect Bell states and that
their particle detectors are perfectly efficient. If these conditions are
not met, our proofs lose their ''all or nothing'' character and can be
rescued only by devising inequalities that are satisfied by local realism
but violated by quantum mechanics (and experiment). We now exhibit one such
inequality. Suppose Alice and Bob share $n$ EPR singlets, with Alice
possessing one member of each pair and Bob the other. Consider the operator $%
B=B_{1}B_{2}...B_{n}$, where $B_{i}=\sigma ^{i}\cdot \hat{a}(\sigma
^{i^{^{\prime }}}\cdot \hat{b}+\sigma ^{i^{^{\prime }}}\cdot \hat{b}^{\prime
})+\sigma ^{i}\cdot \hat{a}^{\prime }(\sigma ^{i^{^{\prime }}}\cdot \hat{b}%
-\sigma ^{i^{^{\prime }}}\cdot \hat{b}^{^{\prime }})$ is the usual CHSH
operator[18] for the $i-$th singlet shared by Alice and Bob, with the unit
vectors $\hat{a},\hat{a}^{\prime },\hat{b},\hat{b}^{^{\prime }}$ being
chosen so as to make the expectation value of $B_{i}$ in the singlet state
achieve its maximal value of $2\sqrt{2}.$ Then, on taking the expectation
value of $B$ in a direct product of $n$ singlets one finds the value $(2%
\sqrt{2})^{n}$, which is to be contrasted with the maximal value of $2^{n}$
yielded by local realism. One therefore finds that the gulf between quantum
mechanics and local realism grows exponentially with the number, $n$, of
singlets considered, which parallels Mermin's finding[19] for the $n-$%
particle GHZ state. It should be added that this Bell inequality does not
rest upon a BKS proof, as was the case with our earlier inequality-free
proofs. The same technique of ''amplification'' used here can be applied to
qudits (i.e. higher spin particles) as well to produce a larger gulf between
the predictions of local realism and quantum mechanics.

The reader may wonder whether the multi-observer GHZ proof can be reduced to
a two-observer proof of the sort discussed here by giving one particle to
one observer and the other $n-1$ particles to a second observer, who is
situated nonlocally with respect to the first. However this will not work
for the following reason. Consider, for simplicity, a three-particle GHZ
state and suppose that particle $1$ is given to Alice and particles $2$ and $%
3$ to Bob. Following Mermin's procedure in [4] Alice and Bob would each
measure the $x-$ and $y-$ component of spin of their particles and they
would also jointly measure the four nonlocal observables 
\begin{equation}
O_{1}=\sigma _{x}^{1}\sigma _{y}^{2}\sigma _{y}^{3}(+1),\text{ \ \ }%
O_{2}=\sigma _{y}^{1}\sigma _{x}^{2}\sigma _{y}^{3}(+1),\text{ \ \ \ }%
O_{3}=\sigma _{y}^{1}\sigma _{y}^{2}\sigma _{x}^{3}(+1)\text{\ \ \ and \ \ }%
O_{4}=\sigma _{x}^{1}\sigma _{x}^{2}\sigma _{x}^{3}(-1),
\end{equation}
where the eigenvalue of each is indicated after it in parentheses. The GHZ
proof works by showing that the product of the values of the observables $%
O_{1},O_{2},O_{3}$ and $O_{4}$ must be $+1$ (in contradiction to what is
implied by (8)) because each of the individual spin components is an element
of reality and occurs twice in the product $O_{1}O_{2}O_{3}O_{4}$. However,
if particles $2$ and $3$ are possessed by a single observer, only the
products of their spin components $\sigma _{y}^{2}\sigma _{y}^{3},\sigma
_{x}^{2}\sigma _{y}^{3},\sigma _{y}^{2}\sigma _{x}^{3}$ and $\sigma
_{x}^{2}\sigma _{x}^{3}$ are elements of reality and, because each occurs
only once in the product $O_{1}O_{2}O_{3}O_{4}$, one can no longer conclude
that this product has to have the value $+1$. One thus sees that the
contradiction on which the three-particle GHZ proof is based disappears if
the particles are shared by just two observers.

To conclude, we have presented a heirarchy of joint BKS-Bell proofs based on 
$n$ identical Bell states shared symmetrically between two observers. Our
proofs illustrate the close relationship between the two foundational
theorems of quantum mechanics and show particularly how the weaker (BKS)
theorem can serve as a catalyst in the proof of the stronger (Bell) one.
Finally, we have derived a generalized CHSH inequality for $n$ Bell states
shared symmetrically between two observers and shown that quantum mechanics
violates this inequality by an amount that increases exponentially with
increasing $n$.

\textbf{Acknowledgement. }I would like to thank David Mermin and Ad\'{a}n
Cabello for several valuable comments on an earlier draft of this paper and
Vlad Babau for some discussions during the early stages of this work.

\bigskip

\bigskip

Fig.1. A 3 x 3 array of observables for a pair of qubits used by Mermin
(ref.9) to prove the Bell-Kochen-Specker theorem.

\bigskip

\bigskip


\begin{thebibliography}{99}
\bibitem{11}  See, however, Mermin's paper in ref.9 for a thought provoking
theoretical argument in support of noncontextuality.
\bibitem{12}  A more detailed explanation of this assertion is as follows.
When Alice measures a particular observable on her qubits, she collapses
them into the two-dimensional subspace associated with a particular
eigenvalue $(+1$ or $-1)$ of that observable. The correlations in state (1)
then dictate that Bob's qubits collapse into the same two-dimensional
subspace of their Hilbert space. If Bob subsequently measures the same
observable as Alice, either alone or in combination with any other
observables that commute with it, his qubits remain within the selected
two-dimensional subspace and he definitely obtains the same eigenvalue as
Alice for the common observable measured.
\end{thebibliography}
\end{document}